\documentclass[article,showpacs,preprintnumbers,amsmath,amssymb]{revtex4}

\usepackage{graphicx}
\usepackage{revsymb}
\usepackage{dcolumn}
\usepackage{bm}

\begin{document}
\title{Time-dependent unitary perturbation theory \\ for intense laser driven molecular orientation}

\author{D. Sugny}
\email{dominique.sugny@ppm.u-psud.fr}
\author{A. Keller}
\author{O. Atabek}
\affiliation{Laboratoire de Photophysique Mol\'{e}culaire du CNRS, Universit\'{e} Paris-Sud, B\^{a}t. 210 - Campus d'Orsay, 91405 Orsay Cedex, France}
\author{D. Daems}
\email{ddaems@ulb.ac.be}
\affiliation{Center for Nonlinear Phenomena and Complex Systems, Universit\'{e} Libre de Bruxelles, CP 231, 1050 Brussels, Belgium}
\author{S. Gu\'erin}
\email{sguerin@u-bourgogne.fr}
\author{H. R. Jauslin}
\affiliation{Laboratoire de Physique de l'Universit\'e de Bourgogne, UMR CNRS 5027, BP 47870, 21078 Dijon, France}
\begin{abstract}
We apply a time-dependent perturbation theory based on unitary transformations combined with averaging techniques,
on molecular orientation dynamics by ultrashort pulses.
We test the validity and the accuracy of this approach on LiCl described within a rigid-rotor model
and find that it is more accurate  than other approximations.
Furthermore, it is shown that a noticeable orientation can be achieved for experimentally standard short laser pulses of zero time average.
In this case, we determine the dynamically relevant parameters by using the perturbative propagator, that is derived from this scheme,
and we investigate the temperature effects on the molecular orientation dynamics.
\end{abstract}
\maketitle

\section{Introduction}

Molecular alignment and orientation induced by an intense laser field continue to be a challenge to both experiment and theory.
On the experimental side, it has already been shown that these processes have a large variety of applications extending from chemical
reactivity to nanoscale design \cite{revnew,exp1,exp2,exp3,exp4,seid3,seid4,ban2}. From the theoretical point of view,
several basic mechanisms aiming at limiting the angular range of the molecular rotational motion have been derived in order
to improve their control \cite{atabek,ban1,ali1,ali2,ali3,dion3}.
In this context, one of the most efficient mechanism is the "kick mechanism" \cite{seid1,seid2,dion,henr,kickn1,kickn2,dion2},
which consists in a sudden angular momentum transfer to the molecule by a so-called half-cycle pulse, which features
a large asymmetry in the magnitude of the positive and negative peak values. The positive part, which is of interest
in the process, is of short duration as compared to the free rotational period of the molecule, and the negative part
is a weak and long tail that has a limited effect on the dynamics and can be neglected as a first approximation \cite{dion}.
It is interesting to note that,
no matter the intensity of such a field, a small parameter can be introduced by rescaling the time in the Schr\"{o}dinger equation
and that perturbative methods can therefore be applied with respect to this parameter \cite{daems1}.

In other respects, it has recently been shown that time-dependent problems generated by short pulses
can be treated by a time-dependent unitary perturbation theory (TDUPT) \cite{daems1,daems2,sche},
 which is the time-dependent version of the perturbation theory using averaging techniques
\cite{vanvleck,cptn1,cptn2,sche2,jaus,lomb2,revue}. Analogous in its spirit to the interaction picture, this theory consists of a series of unitary transformations,
which are aimed at rewriting the evolution operator as a product of other propagators.
Moreover, order by order (Van Vleck) and superexponential (Kolmogorov-Arnold-Moser) algorithms can be derived.
A detailed calculation scheme is provided in recent papers \cite{daems1,daems2} and its efficiency has been shown for a two-level Hamiltonian.

The main purpose of the present article is to extend this study and to test the efficiency of the procedure
as applied to the orientation dynamics of a diatomic molecule driven by an ultra-short electromagnetic field as a prototype.
We demonstrate that the first order propagator can reproduce the time evolution in satisfactory agreement with exact results.
We also show that an appropriate choice for a free parameter available in the TDUPT can help improving the accuracy
of the approximate propagator.
Finally, we use the resulting perturbative propagator to investigate the postpulse molecular dynamics
and to reveal its principal features.
In this paper, we consider short pulses with no asymmetry in their temporal shape,
which means that the time average of the electromagnetic field over this short duration is zero.
Such pulses are actually those which are experimentally achievable \cite{pulse1,pulse2}.
As the standard sudden impact approximation \cite{henr} is not a good starting point in this case, we construct a perturbative propagator using TDUPT.
This enables us to determine the relevant parameters which control the orientation dynamics and to show that noticeable orientation
can be obtained in this case. We also investigate the robustness with the temperature of such a mechanism and we remark that
an efficient orientation may be reached for temperatures not exceeding 5 K for the LiCl molecule (which represents
ten $J$ states mainly initially populated).

The remainder of this article is organized as follows : We outline the principles of TDUPT in Sec. \ref{sec2},
paying special attention to the flexibility of the method. Starting from the expression of the Hamiltonian
describing the molecule LiCl (within a rigid-rotor approximation) interacting with a linearly polarized laser pulse,
we derive several propagators from TDUPT and compare them to the exact (numerical) one in Sec. \ref{molor}.
The connections to other perturbation methods, like the Magnus expansion or the sudden impact approximation, are also presented.
Section \ref{oridyn} is devoted to a discussion of the parameters which control the orientation dynamics for experimentally
available short pulses. For a temperature $T=0\, {\textrm K}$, noticeable orientation lasting over 1 ps
(or approximately one tenth of the rotational period) is obtained in the case of a strictly zero time averaged pulse.
The results are also presented at $T=5\, {\textrm K}$ and in spite of temperature effects which tend to decrease the orientation,
it is shown that an efficient orientation can be obtained, but for shorter durations as compared to those observed at $T=0\, {\textrm K}$.

\section{\label{sec2} Time-dependent unitary perturbation theory}

This section recalls the principles of TDUPT as constructed in \cite{daems1,daems2}. The presentation followed here highlights the analogy of this procedure with the spirit of the interaction picture.
We restrict ourselves to the first order, which we shall consider explicitly in the subsequent sections. To first order, it is worth noting that the general scheme described in this section is common to a time-dependent version of the Van Vleck and the time-dependent KAM perturbation procedures. The two methods differ however at higher orders. The reader is referred to  refs. \cite{daems1,daems2} for details.

\subsection{\label{pert} Description of the perturbation procedure}

Let $H(t)$ be a time-dependent Hamiltonian, and $U_H(t,t_i)$ the corresponding evolution operator. Using atomic units ($\hbar=1$), $U_H(t,t_i)$ is the solution of the Schr\"{o}dinger equation 
\begin{equation}\label{cpt1}
i\frac{\partial}{\partial t}U_H(t,t_i)=H(t)U_H(t,t_i),
\end{equation}
with the initial condition 
\begin{equation}\label{cpt2}
U_H(t_i,t_i)={\textbf 1} ,
\end{equation}
where ${\textbf 1}$ is the identity operator. In order to solve Eq. (\ref{cpt1}) by a perturbative scheme, we assume that a small parameter $\varepsilon$ can be introduced in the following decomposition of the Hamiltonian $H$ 
\begin{equation}\label{cpt3}
H(t)=H_0(t)+\varepsilon V_1(t)  ,
\end{equation}
where $H_0$ is the unperturbed Hamiltonian and $V_1$ the perturbation.
We further assume that the evolution operator $U_{H_0}$ of the time-dependent Hamiltonian $H_0$ is known.\\
The first step of the procedure consists in rewriting the propagator $U_H(t,t_i)$ in the interaction picture.
 Actually, one can introduce an additional parameter $t_p$ (the standard interaction picture being obtained for $t_p=0$)
\cite{henr,mess} 
\begin{equation}\label{cpt4}
U_H(t,t_i)=U_{H_0}(t,t_p)U_{H_1}(t,t_i;t_p)U_{H_0}(t_p,t_i)  .
\end{equation}
It turns out that the truncation of TDUPT at any order is strictly independent of $t_p$ \cite{daems2}.
Hence, throughout the paper we set $t_p=t_i$.
In the interaction picture, Eq. (\ref{cpt1}) reads
\begin{equation}\label{cpt5}
i\frac{\partial}{\partial t}U_{H_1}(t,t_i)=\varepsilon H_1(t)U_{H_1}(t,t_i)  ,
\end{equation}
where the Hamiltonian $H_1$ is defined as
\begin{equation}\label{cpt5a}
H_1(t)=U_{H_0}^{\dagger}(t,t_i)V_1(t)U_{H_0}(t,t_i)  .
\end{equation}
Generally, the Hamiltonian $H_1$ cannot in turn be partionned in a form similar to Eq. (\ref{cpt3}) with higher powers of $\varepsilon$ preventing thus the iteration of the procedure. However, a time-dependent version of unitary perturbation theory can be derived allowing the iteration through a series of unitary transformations.

Following this scheme, the next step of the iteration consists in finding a unitary transformation : $T_1(t;t_2)=e^{-i\varepsilon W_1(t;t_2)}$, where $W_1$ is self-adjoint, such that the resulting Hamiltonian $\tilde H_1$, which will be explicitly defined below, can be decomposed in the form 
\begin{equation}\label{cpt6}
\varepsilon\tilde H_1(t)=\varepsilon D_1(t)+\varepsilon^2 V_2(t)  .
\end{equation}
In this expression, $D_1$ is such that its evolution operator $U_{D_1}$ can be easily calculated, and $\varepsilon^2 V_2$ contains no terms of order lower than 2 in  $\varepsilon$. Application of $T_1$ to the evolution operator $U_{H_1}$, according to the relation 
\begin{equation}\label{cpt7}
U_{H_1}(t,t_i)=T_1(t;t_2)U_{\tilde H_1}(t,t_i;t_2)T_1^{\dagger}(t_i;t_2)  ,
\end{equation}
leads to the following expression for the Schr\"{o}dinger equation [Eq. (\ref{cpt5})] 
\begin{equation}\label{cpt8}
i\frac{\partial}{\partial t}U_{\tilde H_1}(t,t_i)=[\varepsilon e^{i\varepsilon W_1(t)}H_1(t)e^{-i\varepsilon W_1(t)}+i\frac{\partial e^{i\varepsilon W_1(t)}}{\partial t}e^{-i\varepsilon W_1(t)}]U_{\tilde H_1}(t,t_i)  .
\end{equation}
For the sake of notation we omit the dependence on $t_2$ of $T_1$, $W_1$ and $U_{\tilde H_1}$.
The Hamiltonian $\tilde H_1$ introduced in Eq. (\ref{cpt6}) is accordingly defined as 
\begin{equation}\label{cpt8a}
\varepsilon\tilde H_1(t)=\varepsilon e^{i\varepsilon W_1(t)}H_1(t)e^{-i\varepsilon W_1(t)}+i\frac{\partial e^{i\varepsilon W_1(t)}}{\partial t}e^{-i\varepsilon W_1(t)}  .
\end{equation}
Expanding Eq. (\ref{cpt8a}) to first order in $\varepsilon$, and taking into account Eq. (\ref{cpt6}) one obtains the time-dependent generalization of the  cohomological equation \cite{jaus} relating  $W_1$ to $D_1$ and $H_1$ 
\begin{equation}\label{cpt9}
\frac{\partial W_1}{\partial t} = H_1(t)-D_1(t) .
\end{equation}
The general solution to this equation reads 
\begin{equation}\label{cpt10}
W_1(t;t_2)=\int_{t_2}^t du [H_1(u)-D_1(u)] ,
\end{equation}
where $t_2$ is a free parameter. The important point to realize here is that this solution is not unique :
 The perturbation procedure offers the flexibility to control what part of $H_1$ should be retained in $W_1$ by a judicious choice of $D_1$ \cite{daems2}.
We consider explicitly several such choices in the next section.

To summarize, with the help of the unitary transformation $T_1$ we have shown that the evolution operator $U_{\tilde H_1}$ satisfies the following Schr\"{o}dinger equation 
\begin{equation}\label{cpt11}
i\frac{\partial}{\partial t}U_{\tilde H_1}(t,t_i)=\left\{\varepsilon D_1(t)+\varepsilon ^2 V_2(t)\right\}U_{\tilde H_1} (t,t_i) ,
\end{equation}
formally similar to Eq (\ref{cpt5}), but with a Hamiltonian $\tilde H_1$ [given by Eq. (\ref{cpt6})] decomposed as $H$ [Eq. (\ref{cpt3})]. This allows for the iteration of the first step of the procedure. Writing $U_{\tilde H_1}$ in the new interaction picture, we obtain 
\begin{equation}\label{cpt12}
U_{\tilde H_1}(t,t_i)=U_{D_1}(t,t_q)U_{H_2}(t,t_i;t_q)U_{D_1}(t_q,t_i) ,
\end{equation}
where $t_q$ is arbitrary, and set to $t_i$ as, again, the final result is independent of its value. $U_{H_2}$ is the solution of the following equation similar to Eq. (\ref{cpt5}) 
\begin{eqnarray}\label{cpt13}
i\frac{\partial}{\partial t}U_{H_2}(t,t_i)&=&\varepsilon ^2 H_2(t) U_{H_2}(t,t_i) \nonumber \\&=&\varepsilon ^2 U_{D_1}^{\dagger} (t,t_q)V_2(t) U_{D_1}(t,t_q)U_{H_2}(t,t_i) .
\end{eqnarray}
Up to this point, the procedure involves no approximation as it is just a sequence of unitary transformations. The first order approximation consists in replacing the evolution operator $U_{H_2} $ by the identity since it is associated with an Hamiltonian of order $\varepsilon^2$ and is therefore equivalent  to  $U_{H_2}={\textbf 1}+O(\varepsilon ^2)$.
Combining Eqs. (\ref{cpt4}), (\ref{cpt7}) and (\ref{cpt12}), we finally obtain 
\begin{equation}\label{cpt14}
U_H(t,t_i)=U_{H_0}(t,t_i)T_1(t)U_{D_1}(t,t_i)T_1^{\dagger}(t_i)+O(\varepsilon ^2)  .
\end{equation}
Equation (\ref{cpt14}) together with the various choices of $D_1$ considered in the next section will be applied for studying molecular orientation through Sec. \ref{molor}.

\subsection{Choices for improving the accuracy}
As it has been mentioned in the previous section, the principal point one can play with to favor the convergence of the perturbation procedure is the control of the terms of $H_1$ to be kept in $W_1$ through the choice of $D_1$. Three possibilities are considered : \begin{itemize}
\item[i)] For a finite duration, examination of Eq. (\ref{cpt10}) shows that no term must necessarily be kept in $D_1$ to ensure a finite operator $W_1$. The simplest perturbative propagator is therefore obtained by taking $D_1=0$. Referring to Eq. (\ref{cpt14}), one is thus left with an evolution operator, $U^M$, of the form 
\begin{equation}\label{cpt15}
U^M(t,t_i)=U_{H_0}(t,t_i)\exp\left(i\varepsilon\int_{t_i}^t du H_1(u)\right)  .
\end{equation}
In this case, we remark that the first order propagator $U^M$ is completely equivalent to the usual first order propagator obtained with the Magnus formula for the Hamiltonian $H_1$ \cite{magnus}. The general relation between Magnus expansion and TDUPT has been derived in \cite{daems2}.
\item[ii)] Although not mandatory, secular terms, noted $S_1(t)$, can be taken into account in the definition of the operator $D_1$. In the next section, we will see through numerical tests the role of such terms in the convergence of the perturbative propagator. Using an averaging method for time-dependent Hamiltonians \cite{daems1,sche}, $S_1$ can be defined by the following expression 
\begin{equation}\label{cpt16}
S_1=\lim_{T\to +\infty}\frac{1}{T}\int_{t-T}^t du H_1(u) .
\end{equation}
If $H_1$ is an operator constant in time up to a time $t_i$ and with an arbitrary uniformly bounded dependence on time for $t>t_i$, then it has been shown \cite{daems1} that $S_1=H_1(t_i)$.
The choice $D_1=S_1$ allows one to determine the corresponding propagator as 
\begin{equation}\label{cpt16aa}
U_{D_1}(t,t_i)=e^{-i\varepsilon H_1(t_i)(t-t_i)}  .
\end{equation}
\item[iii)] Owing to the reduction of the error upon addition of secular terms in the definition of the operator $D_1$, the question that naturally arises concerns the reduction of the error upon inclusion of other terms. Keeping in mind that the operator $D_1$ can be a solution of our problem as long as one is able to calculate the propagator $U_{D_1}$,  we see that this second step is, however, far from being as obvious as the first one. Nevertheless, another simple and efficient solution consists in choosing the operator $D_1$ such that \cite{daems2} 
\begin{equation}\label{cpt16a}
D_1=H_1(t_1)  ,
\end{equation}
where $t_1$ is a free parameter. Moreover, since $D_1$ does not depend on time, the calculation of $U_{D_1}$ turns out to be very simple.
We shall see that $t_1$ plays a significant role which can enhance the accuracy of the procedure by several orders of magnitude.
This possibility, already present at the first order considered here, stems from the fact that to preserve unitarity we do not expand
the exponentials, retaining thereby terms of higher orders.
\end{itemize}

Having determined the operator $D_1$, we can calculate the generator $W_1$ from Eq. (\ref{cpt10}) and apply the general scheme of the procedure. However, we shall point out an additional adjustment possibility of this approach which is of crucial importance with the intention of improving the accuracy of the approximate propagator. This possibility consists in writing the Hamiltonian $H_1$ defined by Eq. (\ref{cpt5}) in the following form 
\begin{equation}\label{cpt17}
\varepsilon H_1=\varepsilon D_1+\varepsilon (H_1-D_1)  .
\end{equation}
The first step of the procedure can then be applied to the Hamiltonian $\varepsilon H_1$, where $H_0$ is replaced by $\varepsilon D_1$ and $V_1$ by $H_1-D_1$. In this interaction picture, one can rewrite the propagator $U_{H_1}$ as 
\begin{equation}\label{cpt18}
U_{H_1}(t,t_i)=U_{D_1}(t,t_0)U_{\widehat H_1}(t,t_i;t_0)U_{D_1}(t_0,t_i) ,
\end{equation}
where $t_0$ is a free parameter (set to $t_i$ from now as the final result is $t_0$-independent) and $\widehat H_1$ the resulting Hamiltonian, which plays the role of the Hamiltonian $H_1$ in the general procedure described in Sec. \ref{pert}. However, as $D_1$ has already been taken into account in the previous interaction picture, it is important to realize that the new operator $\widehat D_1$ calculated from the Hamiltonian $\widehat H_1$ is generally taken as 0. The propagator obtained at first order in $\varepsilon$ with $\widehat D_1=0$ is hereafter referred as $U^I$.\\

How these adjustments are best performed for studying the orientation dynamics of diatomic molecules driven by a pulsed laser will be the subject of the next section.

\section{\label{molor} Application to the orientation dynamics of diatomic molecules driven by a pulsed laser}

This section is devoted to the application of TDUPT for studying the orientation dynamics of polar diatomic molecules driven by an electromagnetic field.
\subsection{Description of the model}

We consider a molecule described in a rigid-rotor approximation interacting with a linearly polarized laser pulse. The model Hamiltonian is taken to be 
\begin{equation}\label{ori1}
H=BJ^2-\mu _0 E(t)\cos\theta  ,
\end{equation}
where $J^2$ is the angular momentum operator, $B$ the rotational constant, $\mu _0$ the permanent dipole moment (for the sake of simplicity, the polarizability is neglected) and $E(t)$ the electromagnetic field amplitude. $\theta$ is the angle between the direction of the rotor axis and the polarization vector. The values $\mu _0=7.129\ {\textrm D}$ and $B=0.70652\ \, {\textrm cm^{-1}}$ (value at the Li-Cl equilibrium distance \cite{hb}) are chosen so as to reproduce, at least qualitatively, the principal features of the polar diatomic molecule LiCl \cite{dion}. We also recall that in spherical coordinates $\theta$ (polar angle) and $\phi$ (azimuthal angle), $J^2$ stands for the operator \cite{mess} 
\begin{equation}\label{ori2}
J^2=-\frac{1}{\sin\theta}\frac{\partial}{\partial\theta}\sin\theta\frac{\partial}{\partial\theta}-\frac{1}{(\sin\theta)^2}\frac{\partial ^2}{\partial\phi ^2}  .
\end{equation}
Due to cylindrical symmetry, the projection $m$ of the total angular momentum $j$ on the field polarization axis is a classical constant of motion or a good quantum number.  We shall consider, in particular a pulse shape $E(t)$  of the form
\begin{eqnarray}\label{ori3}
E(t)=\left\{ \begin{array}{ll}
E_0\sin ^2(\pi \frac{t}{\delta})\sin(2\pi f\frac{t}{\delta}) &\textrm{if} \quad 0\leq t\leq \delta \\
 0 & \textrm{elsewhere},\\
\end{array} \right.  
\end{eqnarray}
where $E_0$ is the peak amplitude of the pulse, $f/\delta$ its frequency and $\delta$ its duration. Note that this function has been commonly used in the literature to describe the kick mechanism
 \cite{atabek,dion,henr}. The considerations developed below are not restricted to that pulse shape. In Sec. \ref{oridyn}, we shall be dealing with zero time-averaged pulses, which correspond to an integer value of the parameter $f$ entering Eq. (\ref{ori3}). Experimentally, achievable values for $E_0$ and $\delta$ are taken to be $E_0=1.5\cdot 10^6\, {\textrm \ V\cdot cm^{-1}}$ and $\delta=1 \, {\textrm ps}$. This field duration is furthermore smaller by one order of magnitude than typical molecular rotational periods ($1/B\simeq 10\, {\textrm ps}$).
\subsection{Preliminary calculations and derivation of the perturbative evolution operator}
The time-dependent Schr\"{o}dinger equation describing this system is 
\begin{equation}\label{ori4}
i\frac{\partial}{\partial t}U_H(t,t_i)=\left\{B J^2-\mu _0E(t)\cos\theta\right\}U_H(t,t_i)  .
\end{equation}
Its evolution is governed by two characteristic times : the rotational period $T_{{\rm rot}}=1/\left[B \left\langle J^2 \right\rangle\right]$ of the free molecule (where $\left\langle J^2 \right\rangle$ is the mean value of the operator $J^2$) and the pulse duration $\delta$. Moreover, if $\left|\psi(t)\right\rangle$, the wave function of the field-free system at time $t\geq\delta$, is one of the spherical harmonics $\left|j,m\right\rangle$ then this period can be written in the form 
\begin{equation}\label{ori5}
T_{{\rm rot}}=\frac{1}{Bj(j+1)} .
\end{equation}
We remark that $T_{{\rm rot}}$ decreases with the value of the quantum number $j$.\\
In the sudden limit, i.e., for short-pulse duration with respect to the rotational period, a small dimensionless parameter $\varepsilon$ can be introduced as 
\begin{equation}\label{ori6}
\varepsilon=B \delta  ,
\end{equation}
which amounts to be $\varepsilon\simeq 0.1$ with the numerical values of $B$ and $\delta$. From a practical point of view, provided the pulse duration be sufficiently small, it is expected that this formulation allows one to study the dynamics of the system even in the presence of large non-perturbative pulse areas. Such a conclusion true for a two-level system \cite{daems1} has to be considered more carefully here in the sense that higher peak amplitudes induce higher rotational population leading to shorter periods, such that the choice of $\delta$ actually depends on the pulse area.\\

Rescaling the time in the form $\tau =t/\delta$, we obtain for the Schr\"{o}dinger equation (\ref{ori4}) 
\begin{equation}\label{ori7}
i\frac{\partial}{\partial \tau}U_H(\tau,\tau _i)=\left\{-E_r(\tau)\cos\theta+\varepsilon J^2\right\}U_H(\tau,\tau _i)  ,
\end{equation}
where $E_r(\tau)=\mu _0 \delta E(\delta\tau)$. We note $E_{0r}=\mu _0 \delta E_0$ the peak amplitude of $E_r$ with a typical value of $E_{0r}\simeq 30$ and we introduce the following dimensionless time $\tau_i=0$ and $\tau_f=1$. The pulse duration is then $\tau_i\leq\tau\leq\tau_f$. Here, it is also important to realize that the small parameter $\varepsilon$ is independent of the choice of $E_{0r}$ such that the perturbative method in consideration is consistent with very strong fields.

The procedure described in Sec. \ref{pert} is applied to the Hamiltonian involved in Eq. (\ref{ori7}). In this particular case, we can first define the operators $H_0$ and $V_1$ of Eq. (\ref{cpt3}) as 
\begin{eqnarray}\label{ori7a}
\left\{ \begin{array}{ll}
H_0(\tau) &=-E_r(\tau)\cos\theta \\
 V_1 &=J^2 .  
\end{array} \right. 
\end{eqnarray}
hence
\begin{equation}\label{ori7b}
U_{H_0}(\tau,\tau_i)=e^{iA(\tau)\cos\theta}  ,
\end{equation}
where $A(\tau)=\int_{\tau _i}^\tau du  E_r(u)$ is the pulse area in the interval $[\tau_i,\tau]$. Application of the first step of the procedure [Eq. (\ref{cpt5a})] leads to the following expression for $H_1$ 
\begin{equation}\label{ori7c}
H_1(\tau)=e^{-iA(\tau)\cos\theta}J^2e^{iA(\tau)\cos\theta} .
\end{equation}
Using the Campbell-Hausdorf formula \cite{quant1}
\begin{equation}
\label{CH}
e^C B e^{-C}=B+\frac{1}{1!}[C,B]+\frac{1}{2!}[C,[C,B]]+\cdots,
\end{equation}
 and the commutation relations 
\begin{eqnarray}\label{ori9}
[J^2,\cos\theta] & = & 2(\sigma_{\theta}+\cos\theta ) , \\
\lbrack\sigma_{\theta},\cos\theta] & = & \cos^2\theta-{\textbf 1} ,
\end{eqnarray}
where $\sigma_{\theta}\equiv \sin\theta\frac{\partial}{\partial\theta}$, one can rewrite Eq.~(\ref{ori7c})  exactly  as \cite{revue}
\begin{equation}\label{ori8}
H_1(\tau)=J^2+2iA(\tau)(\sigma_{\theta}+\cos\theta)+A^2(\tau)(\textbf{1}-\cos ^2\theta) .
\end{equation}

Finally, to first order in $\varepsilon$, the propagator of Eq. (\ref{cpt14}) is written in the following form for $\tau _i\leq \tau\leq \tau _f$ 
\begin{equation}\label{ori11}
U_H(\tau,\tau _i)=e^{iA(\tau)\cos\theta}T_1(\tau)U_{\varepsilon D_1}(\tau,\tau _i)T_1^{\dagger}(\tau _i)+O(\varepsilon^2) ,
\end{equation}
The choice for the operators $D_1$ and $T_1$ will be explicited in the next section. For $\tau >\tau _f$, the propagation is assumed to be free 
\begin{equation}\label{ori12}
U_H(\tau,\tau _f)=e^{-i\varepsilon (\tau-\tau _f) J^2}.
\end{equation}

\subsection{Numerical tests}
We are now in a position to check the accuracy of the perturbation scheme through the comparison of wave functions obtained by the various TDUPT propagators with those obtained from the accurate full numerical split-operator method \cite{split1,split2}.
The choices described above for $D_1$ and for the decomposition of the Hamiltonian $H_1$, as well as the values of the time parameters $\tau_1$ and $\tau_2$, lead to different evolution operators, the merits of which are numerically checked hereafter. More precisely, for $\tau_i \leq \tau \leq \tau_f$, we quote :
\begin{itemize}
\item[i)] The choice $D_1=0$ which leads to the first order Magnus propagator through the use of Eqs. (\ref{cpt15}), (\ref{ori7b}) and (\ref{ori11}) 
\begin{equation}\label{ori13}
U^M(\tau,\tau _i)=e^{i A(\tau)\cos\theta}\exp\left\{-i\varepsilon\int_{\tau_i}^\tau du  H_1(u)\right\} .
\end{equation}
\item[ii)] The choice $D_1=S_1$ with a secular term $S_1$ [Eq. (\ref{cpt16})] that can easily be evaluated through Eq. (\ref{ori8})  with the result in the case where $A$ vanishes for $\tau\leq \tau_i$ 
\begin{equation}\label{ori14}
S_1= J^2  ,
\end{equation}
leads to the propagator $U^S$ calculated using Eq. (\ref{cpt16aa}) 
\begin{eqnarray}\label{ori15}
U^S(\tau,\tau _i)&=& e^{i A(\tau)\cos\theta}\exp\left\{-i\varepsilon\int_{\tau_2}^\tau du  \left[H_1(u)-J^2\right]\right\}\nonumber \\ &&\times e^{-i\varepsilon (\tau-\tau_i)J^2}\exp\left\{-i\varepsilon\int_{\tau_i}^{\tau_2} du  \left[H_1(u)-J^2\right] \right\}  .
\end{eqnarray}
\item[iii)] The decomposition defined by Eq. (\ref{cpt17}) with $D_1=H_1(\tau_1)$ as in Eq. (\ref{cpt16a})  leading to 
\begin{eqnarray}\label{ori16}
U^I(\tau,\tau _i)&=&e^{iA(\tau)\cos\theta}e^{-i\varepsilon H_1(\tau_1)(\tau-\tau_i)} \nonumber \\
& &\times \exp\left\{-i\varepsilon\int_{\tau_i}^\tau du  e^{i\varepsilon (u-\tau_i)H_1(\tau_1)}\left[H_1(u)-H_1(\tau_1)\right] e^{-i\varepsilon (u-\tau_i)H_1(\tau_1)}\right\}  .
\end{eqnarray}
In the particular case $\tau_1=\tau_i$, one deduces from Eq. (\ref{ori8}) that 
\begin{equation}\label{ori16bis}
H_1(\tau_1=\tau_i)=J^2  .
\end{equation}
\end{itemize}
It is worth mentioning that all these propagators are close to the Magnus one in a sense specified in \cite{daems2}. However, from a pratical point of view, the fact that the operators $J^2$, $\sigma_{\theta}$ and $\cos\theta$ do not commute leads to different propagators.\\
We also recall the form of the propagator $U^{SI}$ that has already been used for this problem in the sudden impact approximation \cite{henr}.
 Rewriting the initial Hamiltonian [Eq. (\ref{ori7})] in the interaction picture involving the operator $J^2$ and neglecting the molecular rotational motion during the pulse, yields at time $\tau_f$ 
\begin{equation}\label{ori16a}
U^{SI}(\tau_f,\tau_i)=e^{-i\varepsilon (\tau_f-\tau_h)J^2}e^{iA(\tau_f)\cos\theta}e^{-i\varepsilon (\tau_h-\tau_i)J^2}  ,
\end{equation}
where $\tau_h$ is a parameter. Numerical tests show that the best choice for this free parameter is $\tau_h=(\tau_f-\tau_i)/2$.
 It is interesting to note that this parameter $\tau_h$ although introduced through an interaction representation  (see Eq. (\ref{cpt4}) and remark below this equation) does affect the final result.
This is due to the fact that in \cite{henr} an additional approximation is made which assumes an impulsive character for the perturbation.
\\

In order to measure the accuracy of these propagators, we define the error $\Delta$ at the end of the pulse as 
\begin{equation}\label{ori17}
\Delta=\Arrowvert \psi(\tau_f)-\psi^{ex}(\tau_f)\Arrowvert^2
\end{equation}
where $\psi^{ex}$ is the exact wave function computed by solving the time-dependent Schr\"{o}dinger equation with the split-operator method and $\psi$ the one obtained by applying the propagator under consideration. In the numerical cases studied below, the initial wave function $\psi(\tau_i)$ is taken as the ground rotational state $\left|j=0,m=0\right\rangle$ of the molecule. Note that this state can be experimentally prepared, for instance, by laser cooling methods \cite{ata2}.
\begin{figure}
\includegraphics[width=1.0\textwidth]{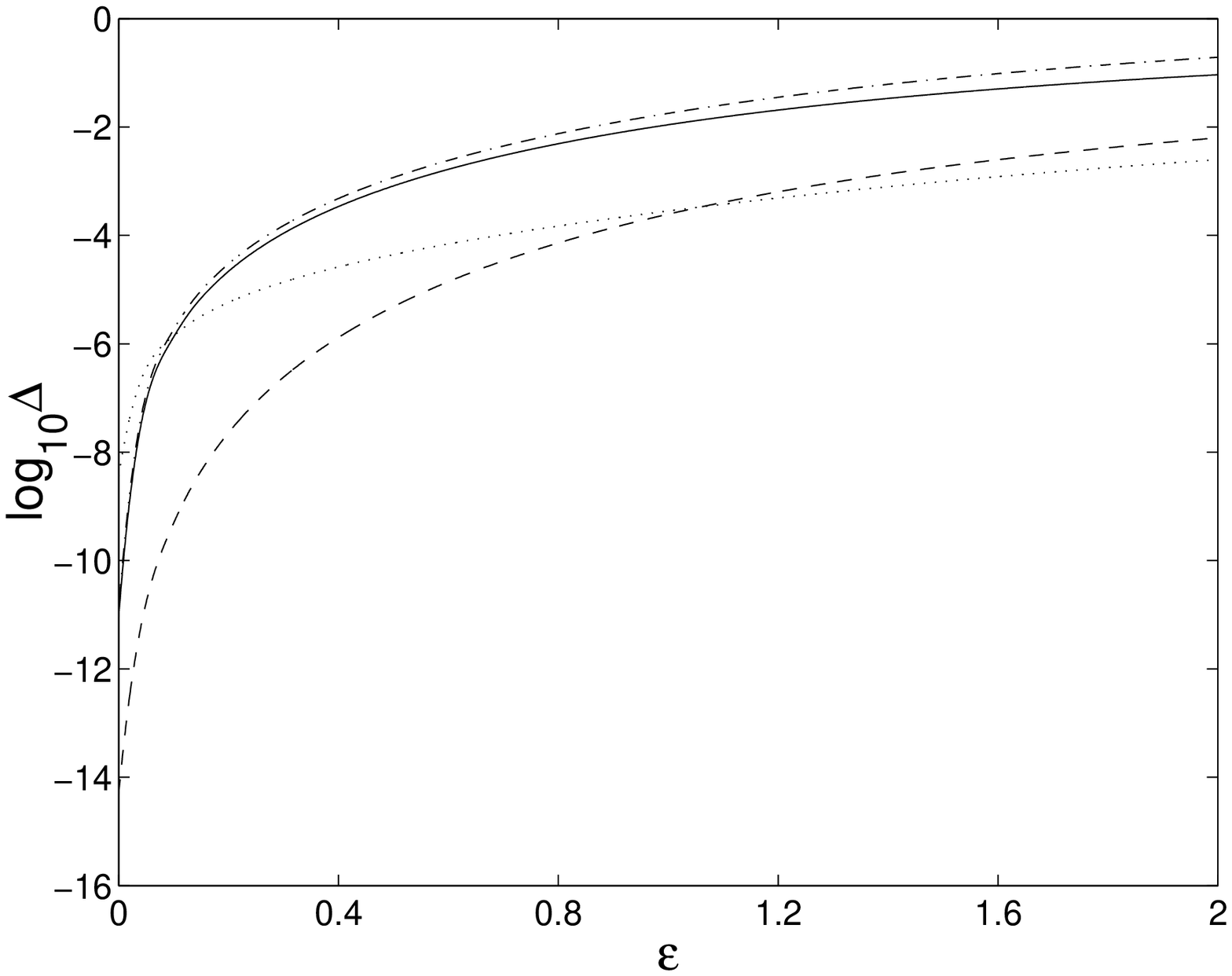}
\caption{\label{comp1}  Common logarithm of the error $\Delta$ [defined in Eq. (\ref{ori17})] as a function of the adimensional} parameter $\varepsilon$ for $E_{0r}=1$ and $f=0.5$. The solid line depicts the error for the propagator $U^M$, the dashed line for $U^I$, the dash-dotted line for $U^S$ and the dotted line represents this error for $U^{SI}$. The free time parameters $\tau_1$ and $\tau_2$ are fixed to 0.
\end{figure}
\begin{figure}
\includegraphics[width=1.0\textwidth]{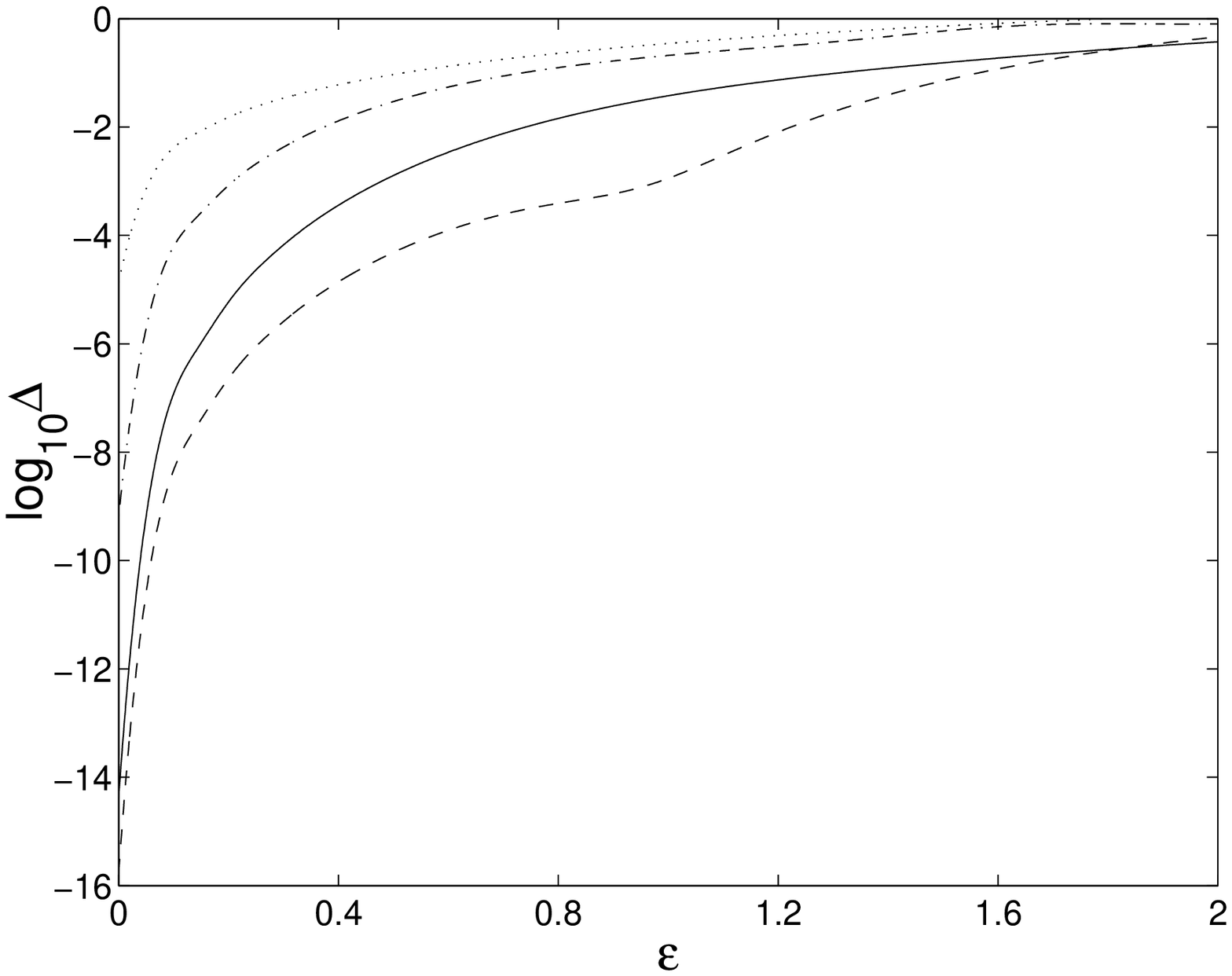}
\caption{\label{comp2} Same as Fig \ref{comp1}, but for $f=2$ and $E_{0r}=20$.}
\end{figure}

\begin{figure}
\includegraphics[width=1.0\textwidth]{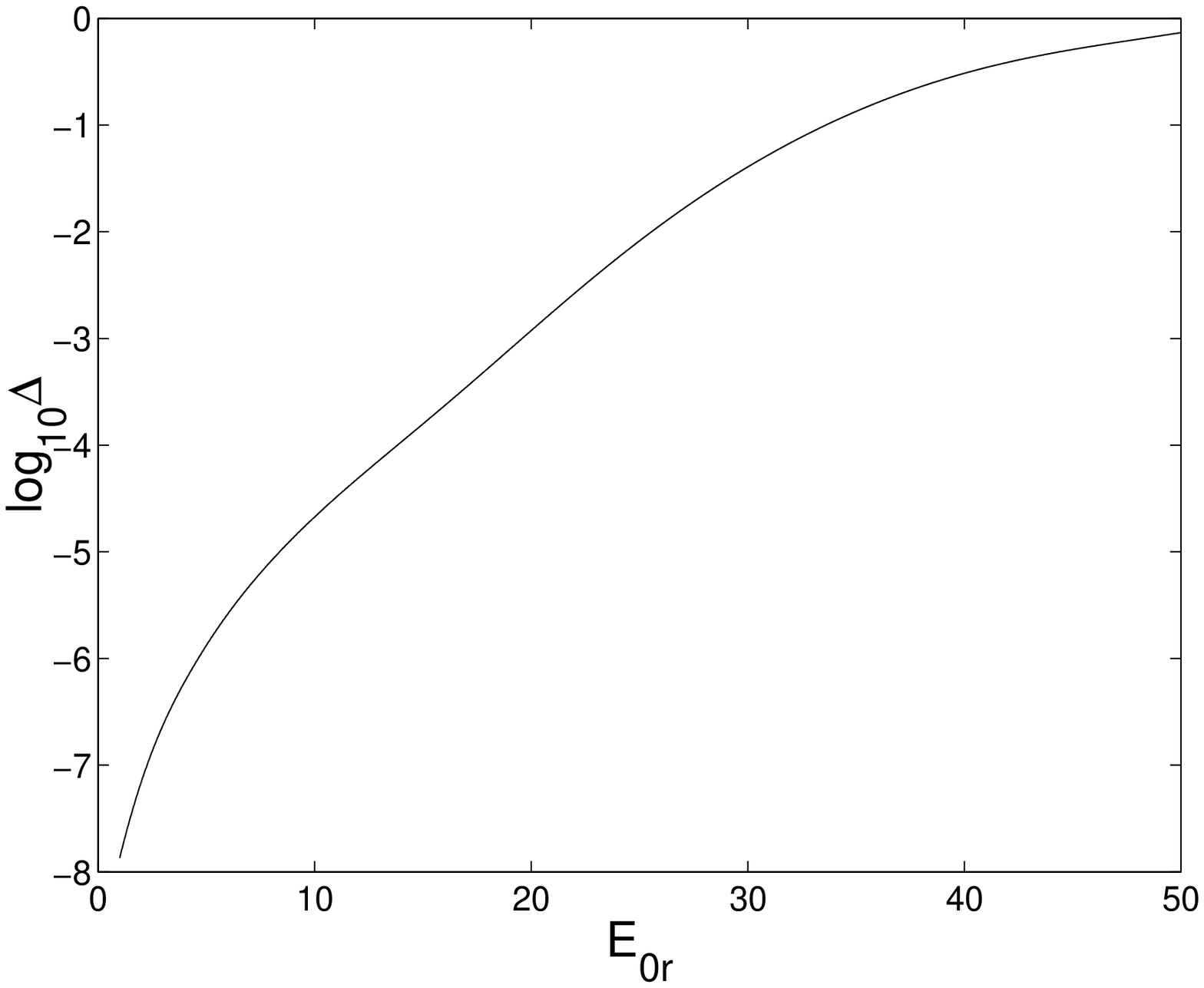}
\caption{\label{amplitude}  Common logarithm of the error $\Delta$  as a function of the adimensional} parameter $E_{0r}$ for the propagator $U^I$ in the case $\varepsilon=1$ and $f=2$. The free time parameter $\tau_1$ is fixed to 0.
\end{figure}

The remarkable accuracy of TDUPT propagators is clearly illustrated in Fig. \ref{comp1} and \ref{comp2}, which display the logarithm of $\Delta$ as a function of the parameter $\varepsilon$ for the following evolution operators $U^M$, $U^{SI}$, $U^S$ and $U^I$. Numerical values are $E_{0r}=1$ and $f=0.5$ in Fig. \ref{comp1} and $E_{0r}=20$ and $f=2$ in Fig. \ref{comp2}. The pulse area is zero when $f$ is an integer, which allows us to consider stronger fields in this case. It is also noted that, for an half-cycle pulse ($f=0.5$), the maximum peak amplitude that can be experimentally achieved is smaller than $E_0=1.5\cdot 10^5\, {\textrm \ V\cdot cm^{-1}}$ \cite{dion}, which corresponds to a parameter $E_{0r}$ of the order of 3. As could be expected, the general trend is that the error decreases as the parameter $\varepsilon$ decreases. In this way, the exact result is reproduced within an accuracy better than $10^{-3}$ when $\varepsilon\leq 0.5$. More unexpectedly, the agreement is still quite good for larger values of $\varepsilon$ (see \cite{daems2} for an explanation). Indeed, the error for $\varepsilon=1$ is as small as $\Delta\simeq 10^{-4}$ for $U^I$ and $f=2$, which is particularly impressive considering that only one iteration is used. Moreover, we remark that the secular terms do not decrease the error of the perturbative propagator $U^S$ in comparison with $U^M$ whereas, with the addition of the last adjustement [Eq. (\ref{cpt17})], $\Delta$ is roughly reduced by, at least, two orders of magnitude for $U^I$. For $\varepsilon\leq 1$, $U^I$ is much more accurate than the sudden impact evolution operator $U^{SI}$. This difference is particularly striking when the pulse area is zero ($f=2$, Fig. \ref{comp2}).
We next analyze the logarithm of $\Delta$ as a function of the parameter $E_{0r}$ for the propagator $U^I$. Such a plot appears in Fig. \ref{amplitude}, where numerical values for $\varepsilon$ and $f$ are taken to be $\varepsilon=1$ and $f=2$. As could be expected, the most salient feature of Fig. \ref{amplitude} is the fact that the error increases with $E_{0r}$. This means that the rotational population is strongly modified and that higher $J$'s have to be taken into account.

The second part of this section concerns the role of the free time parameters $\tau_1$ and $\tau_2$ in the lowering of the error.
First notice that $U^M$ and $U^I$ do not depend on $\tau_2$. For $\tau_2$-dependent TDUPT propagators as $U^S$, it can be shown that this dependence is not dramatic and the best choice is $\tau_2=\tau_i$ \cite{daems2}.
\begin{figure}
\includegraphics[width=1.0\textwidth]{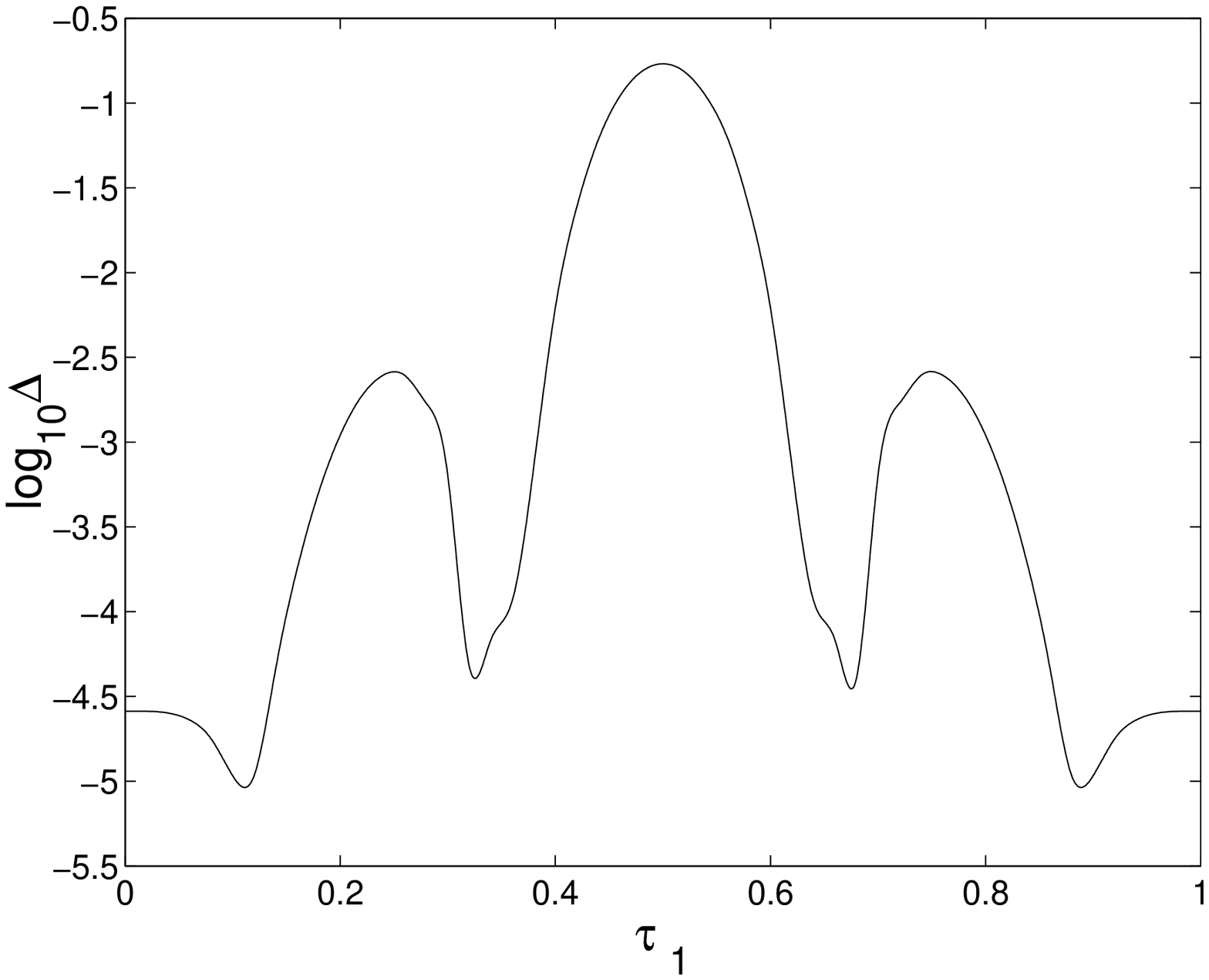}
\caption{\label{time} Common logarithm of the error $\Delta$  as a function of the adimensional} time parameter $\tau_1$ for the propagator $U^I$  in the case $\varepsilon=1$, $f=2$ and $E_{0r}=10$.
\end{figure}

Figure \ref{time} displays the logarithm of $\Delta$ for the propagator $U^I$ as a function of the time parameter $\tau_1$. Numerical values for $\varepsilon$, $f$ and $E_{0r}$ are taken to be $\varepsilon=1$, $f=2$ and $E_{0r}=10$. In this example, it is clear that the best choice for this parameter is $\tau_1\simeq 0.11$ (or $\tau_1\simeq 0.89$). This degree of freedom can be optimized without any, {\textit a priori}, knowledge of the exact solution by locating the minimum with respect to $\tau_1$ of the eigenvalues of an appropriate operator \cite{daems2}.\\

As explained in this section, several choices and attempts can be made to improve the TDUPT propagators. Nevertheless, the best choice of the operator $D_1$ or the best set of time parameters remains dependent upon the context. As long as no dynamical information is lost, the best choice is the one for which the perturbative propagator is as simple as possible.

\section{\label{oridyn} Orientation dynamics by short pulse of zero time-average}
\subsection{Zero rotational temperature}
 We first consider the limit of low rotational temperatures.
Having demonstrated the validity and the accuracy of the perturbative propagator, we now use it to investigate molecular orientation dynamics after the pulse is over. More precisely, we are looking for the relevant parameters of the laser improving the orientation of the molecule. The mean value $\left\langle\cos\theta\right\rangle_{\tau}$ is taken as a quantitative measure of orientation \cite{dion,frid}
\begin{equation} \label{dyn1}
\left\langle \cos\theta\right\rangle_{\tau}=\left\langle\psi(\tau)| \cos\theta| \psi(\tau)\right\rangle .
\end{equation}
It is to be noted that a good orientation is obtained for large absolute values of $\left\langle\cos\theta\right\rangle_{\tau}$ and that the measure used does not take into account the temperature effects \cite{kickn1,kickn2,tempe}, which will be investigated in the second part of this section.
\\

\begin{figure}
\includegraphics[width=1.0\textwidth]{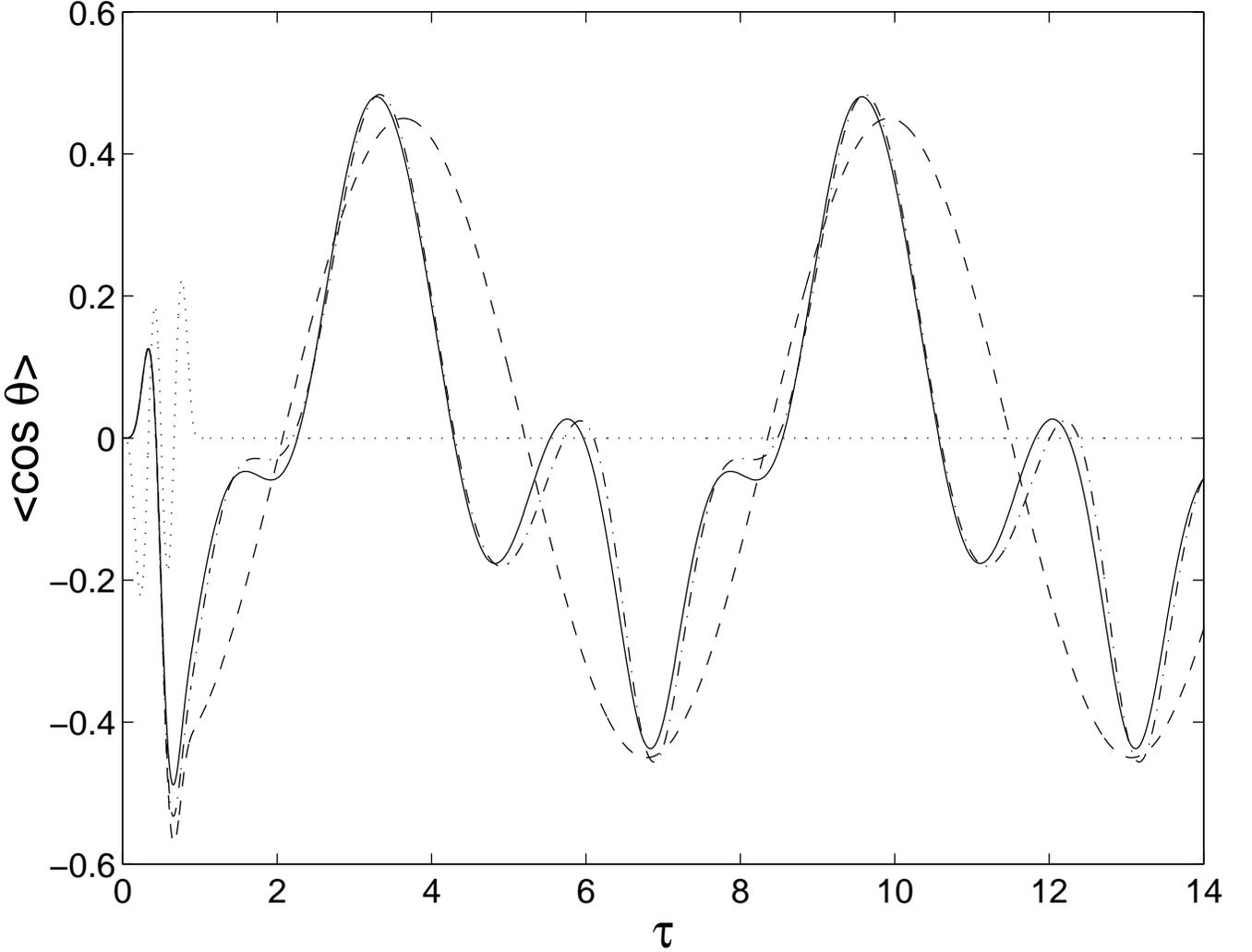}
\caption{\label{cosap}  Mean value $\left\langle\cos \theta \right\rangle_{\tau}$ as a function of the adimensional time $\tau$  computed with the split-operator method (solid line), the sudden impact propagator $U^{SI}$ (dotted line), and the  perturbative propagator $U^I$ according to Eq. (\ref{fullcos}) [dash-dotted line] and Eq. (\ref{cos1}) [dashed line].
Here  $T=0 \, {\textrm K}$ for $\varepsilon=0.5$, $f=2$, $E_{0r}=50$ and the pulse is on for $0\leq\tau\leq 1$.}
\end{figure}

In order to highlight the role of TDUPT in the understanding of molecular orientation, we consider short pulses with symmetrical temporal shape, which means that the time average of the radiative field over this short duration is zero as it has to be for a freely propagating electromagnetic pulse \cite{pulse1}. To our knowledge, it has never been shown that such pulses can be used to obtain a good orientation. Moreover, according to the sudden impact approximation, so far used in the literature to describe the kick mechanism \cite{dion,henr,dion2}, such (zero time averaged) pulses would lead to post-pulse dynamics without any orientation effects.
Indeed, the sudden impact propagator $U^{SI}$ is given by the following expression [Eq. (\ref{ori16a})] for $\tau >\tau_f$ 
\begin{equation} \label{si}
U^{SI}(\tau,\tau_i)=e^{-i\varepsilon (\tau-\tau_h)J^2}e^{iA(\tau_f)\cos\theta}e^{-i\varepsilon (\tau_h-\tau_i)J^2}  .
\end{equation}
If $A(\tau_f)=0$ and $\psi(\tau_i)=\left|j,m\right\rangle$, it is clear that $\left\langle\cos\theta\right\rangle_{\tau}=0$ when the pulse is turned off. This is also seen in Fig. \ref{cosap} which displays $\left\langle\cos\theta\right\rangle_{\tau}$ as a function of time $\tau$.

Using the propagators we have just derived with TDUPT, we calculate $\left\langle\cos\theta\right\rangle_{\tau}$ after the pulse. We consider the most accurate evolution operator $U^I$ for $\tau\geq\tau_f$ and take  $\tau_1=\tau_i$, so that  Eqs. (\ref{ori16}) and (\ref{ori16bis}) yield
\begin{equation} \label{dyn1a}
U^I(\tau,\tau _i)= e^{-i\varepsilon (\tau-\tau_i)J^2}\exp\left\{-i\varepsilon\int_{\tau_i}^{\tau_f} du  e^{i\varepsilon (u-\tau_i)J^2}\left[H_1(u)-J^2\right] e^{-i\varepsilon (u-\tau_i)J^2}\right\}  .
\end{equation}

From Eq.~(\ref{ori8}) one deduces that
\begin{equation}
H_1(u)-J^2=2iA(\tau)(\sigma_{\theta}+\cos\theta)+A^2(\tau)(\textbf{1}-\cos ^2\theta) ,
\end{equation}
where we recall that $\sigma_{\theta} =\sin \theta \partial/ \partial \theta$.

Let us define the following operator which appears in the second exponential of Eq.~(\ref{dyn1a})
\begin{equation}\label{E}
F\equiv i\int_{\tau_i}^{\tau_f} du e^{i\varepsilon (u-\tau_i)J^2}\left\{H_1(u)-J^2\right\} e^{-i\varepsilon (u-\tau_i) J^2} .
\end{equation}
We rewrite accordingly the propagator as
\begin{equation} \label{UIE}
U^I(\tau,\tau _i)=e^{-i\varepsilon (\tau-\tau_i)J^2}e^{-\varepsilon F}.
\end{equation}
By virtue of the action of the operators $J^2$, $\cos\theta$ and $\sigma_{\theta}$ in the basis of the spherical harmonics 
\begin{eqnarray}\label{dyn3a}
J^2\left|j,m\right\rangle&=& j(j+1)\left|j,m\right\rangle, \nonumber\\
\cos\theta \left|j,m\right\rangle&=& c_{j+1 \, m}\left|j+1,m\right\rangle +c_{j\, m}\left|j-1,m\right\rangle, \nonumber\\
\sigma_{\theta} \left|j,m\right\rangle&=& j c_{j+1\, m}\left|j+1,m\right\rangle  -(j+1)c_{j \, m}\left|j-1,m\right\rangle ,
\end{eqnarray}
where $c_{j\, m}=\left[(j-m)(j+m)/(2j-1)(2j+1)\right]^{1/2}$ ,
it follows that $F\left|j,m\right\rangle$ is of the form
\begin{equation}
F\left|j,m\right\rangle=a_{j \, m} \left|j-1,m\right\rangle+b_{j \, m} \left|j+1,m\right\rangle+\alpha_{j \, m} \left|j-2,m\right\rangle+\beta_{j \, m} \left|j+2,m\right\rangle+\gamma_{j \, m} \left|j,m\right\rangle .
\end{equation}
One readily obtains the coefficients
\begin{eqnarray}
a_{j \, m}&=& 2 j c_{j m} \int_{\tau_i}^{\tau_f} du A(u) e^{-2i \varepsilon j (u-\tau_i)} ,\nonumber\\
b_{j \, m}&=& - 2 (j+1) c_{j+1 m} \int_{\tau_i}^{\tau_f} du A(u) e^{2i \varepsilon (j+1) (u-\tau_i)}  ,\nonumber\\
\alpha_{j \, m}&=& - i  c_{j-1 m} c_{j m} \int_{\tau_i}^{\tau_f} du A^2(u) e^{-2i \varepsilon (2j-1) (u-\tau_i)} , \nonumber\\
\beta_{j \, m}&=& - i  c_{j+1 m} c_{j+2 m} \int_{\tau_i}^{\tau_f} du A^2(u) e^{2i \varepsilon (2j+3) (u-\tau_i)}  ,\nonumber\\
\gamma_{j \, m}&=&i(1-c^2_{j m}-c^2_{j+1 m})\int_{\tau_i}^{\tau_f}du   A^2(u)  .
\end{eqnarray}

For a system initially prepared in the state $\left|j,m\right\rangle$
\begin{equation}
\label{ini}
\left|\psi({\tau_i})\right\rangle=\left|j,m\right\rangle,
\end{equation}
we describe the orientation dynamics at time $\tau$ with the mean value $\left\langle\cos\theta\right\rangle_{\tau}$  computed with the propagator $U^I$  according to Eqs.~(\ref{dyn1}) and (\ref{UIE})
\begin{eqnarray} \label{fullcos}
\left\langle \cos\theta\right\rangle_{\tau}
&=&\left\langle j,m\left| e^{\varepsilon F} e^{i\varepsilon (\tau-\tau_i) J^2} \cos \theta e^{-i\varepsilon (\tau-\tau_i) J^2} e^{-\varepsilon F}\right| j,m\right\rangle .
\end{eqnarray}
Using the Campbell-Hausdorff formula, Eq.~(\ref{CH}), we obtain
 \begin{eqnarray}
\left\langle \cos\theta\right\rangle_{\tau}&=&\varepsilon\left\langle j,m\left| [F, e^{i\varepsilon (\tau-\tau_i) J^2} \cos \theta e^{-i\varepsilon (\tau-\tau_i) J^2} ]\right| j,m\right\rangle  +O(\varepsilon^2)\nonumber\\
&=&4\varepsilon(j+1)c_{j+1\,m}^2\int_{\tau_i}^{\tau_f} du A(u)\cos\left(2\varepsilon[j+1][u-\tau+\tau_i]\right)\nonumber \\
&-&4\varepsilon jc_{j\,m}^2\int_{\tau_i}^{\tau_f} du A(u)\cos\left(2\varepsilon j[u-\tau+\tau_i]\right)+O(\varepsilon ^2).
 \label{dyn3d}
\end{eqnarray}
Note that the higher order terms can be neglected as the propagator $U^I$ is constructed using the first order TDUPT.
In other words, at this level of approximation we have already discarded terms of order $\varepsilon^2$.
 
For very low rotational temperatures, the only rotational level initially populated being $\left|j=0,m=0\right\rangle$ \cite{ata2}, one deduces that 
\begin{eqnarray} \label{dyn4}
\left\langle\cos\theta\right\rangle_{\tau}&=&\frac{4}{3}\varepsilon\left\{\cos(2\varepsilon[\tau-\tau_i])\int_{\tau_i}^{\tau_f} du A(u)\cos(2\varepsilon u) +\sin(2\varepsilon[\tau-\tau_i])\int_{\tau_i}^{\tau_f}du  A(u)\sin(2\varepsilon u) \right\}+O(\varepsilon ^2)  .
\end{eqnarray}
Introducing the Fourier transform $\widehat A$ of the pulse area 
\begin{eqnarray} \label{dyn4b}
\widehat A (k)=\frac{1}{\sqrt{2\pi}}\int_{\tau_i}^{\tau_f}du A(u)e^{-iuk}  ,
\end{eqnarray}
we rewrite Eq.~(\ref{dyn3d})  in the form 
\begin{eqnarray} \label{cos1}
\left\langle\cos\theta\right\rangle_{\tau}&=&\frac{4\sqrt{2\pi}}{3}\varepsilon\left|\widehat A(2\varepsilon)\right|\cos\left(2\varepsilon[\tau-\tau_i]+\arg[\widehat A(2\varepsilon)]\right)+O(\varepsilon^2)  .
\end{eqnarray}
In Figure~\ref{cosap} we display the mean values  $\left\langle\cos\theta\right\rangle_{\tau}$ computed with the propagator $U^I$ according to Eq.~(\ref{fullcos}) and according to its first order contribution as given in Eq.~(\ref{cos1}).
The mean values obtained with the sudden impact propagator of Eq.~(\ref{si}) and purely numerically with the split-operator method are also depicted.
It is seen that the first order expression of Eq.~(\ref{cos1}) brings a significant improvement with respect to the standard sudden impact approach which predicts  no post-pulse orientation effect in the case of short pulses of zero time-average considered here.

It is worth noting that $2\varepsilon$ is the rescaled frequency between the first two rotational levels and that the leading revival structures \cite{seid1} are well described by this approximation up to large values of $\varepsilon$, at least from a qualitative point of view. The other fundamental frequencies are neglected to first order in $\varepsilon$, which physically means that the corresponding levels with higher $j$'s are not strongly populated in comparison with the first two rotational states. From Eq. (\ref{cos1}) one deduces that the range of the post-pulse orientation mainly depends on $\varepsilon$ and $|\widehat A(2\varepsilon)|$ which are the two dynamically relevant parameters one can play with to control $\left\langle\cos\theta\right\rangle_{\tau}$. 

Moreover, thanks to a careful choice of the laser parameters which will be discussed below, it is shown in Fig. \ref{cosap} that a noticeable orientation can be achieved with zero time averaged pulses. Indeed, in this example the value $\left|\left\langle\cos\theta\right\rangle_{\tau}\right|\simeq 0.5$ is reached and lasts for a time larger than 1 ps. To our knowledge, it is the first time that such an orientation is observed in this case.\\
We next analyze the choice of the different parameters. For $\varepsilon$ sufficiently small,  $|\widehat A(2\varepsilon)|$ is well approximated by $\int_{\tau_i}^{\tau_f} du A(u)$.
It is instructive to consider the case where the pulse shape is given by Eq. (\ref{ori3}) for which we see how this latter quantity  depends on the frequency and the peak amplitude 
\begin{eqnarray} \label{dyn5}
\int_{\tau_i}^{\tau_f} du A(u)=\left\{\begin{array}{ll}
E_{0r}\frac{3}{16\pi}(\tau_f-\tau_i)&\textrm{if $f=1$}\\ \\
E_{0r}\frac{1}{4\pi}\frac{-1}{f(f^2-1)}(\tau_f-\tau_i)&\textrm{if $f>1$} .
\end{array}\right.  
\end{eqnarray}
It follows from Eqs. (\ref{cos1}) and (\ref{dyn5}) that the orientation decreases as the frequency $f$ increases.
In the high-frequency regime, where $f\gg 1$, we see that no orientation can be obtained as has already been shown using a high-frequency Floquet approach \cite{arne}. On the other hand, in the range of validity of Eq. (\ref{cos1}), it is important to realize that the increase of $\varepsilon$ and $E_{0r}$ involves a better orientation which, however, lasts for shorter durations (the period of the motion is equal to $\pi/\varepsilon$). Experimentally, this point is of crucial importance because the possibility to perform, for instance, stereodynamically sensitive chemical reactions by using this orientation depends on this period of time. For practical purposes, this duration has to be larger than 1 or 2 ps. Finally, numerical values for $\varepsilon$, $E_{0r}$ and $f$, which fulfill these conditions, are taken to be $\varepsilon=0.5$, $E_{0r}=50$ and $f=2$.
We stress that the conclusions on the choice of parameters are general, and not restricted to the particular case of molecule (LiCl) considered as an illustration.\\

\subsection{Finite rotational temperature}
The results presented above were based on the assumption that the rotational temperature $T$ was zero. We now investigate the temperature effects on molecular orientation. In this case, a thermal average over the rotational levels has to be taken into account. The quantitative measure of the orientation is then given by  \cite{kickn1,kickn2,tempe} 
\begin{equation} \label{dyn6}
\ll\cos\theta\gg_{\tau}=\frac{1}{Q}\sum_j\exp\left(\frac{-Bj[j+1]}{k_BT}\right)\sum_{m=-j}^{j}\left\langle\cos\theta\right\rangle_{\tau} ,
\end{equation}
where $k_B$ is the Boltzmann constant and $Q$ the partition function 
\begin{equation} \label{dyn7}
Q=\sum_j(2j+1)\exp\left(\frac{-Bj[j+1]}{k_BT}\right)  .
\end{equation}

The result for LiCl molecule under the effect of the previous pulse is shown in Fig. \ref{temp}, which displays the thermally averaged mean value $\ll\cos \theta \gg_{\tau}$ as a function of the time $\tau$ for the rotational temperature $T=5 \, {\textrm K}$.

We first note a decrease of the orientation with increasing temperature, this point has already been mentionned in previous studies \cite{kickn1,kickn2,tempe}. Moreover, as for the study at $T=0 \, {\textrm K}$, an approximate analytical formula can be derived for $\ll\cos\theta\gg_{\tau}$. Using Eq. (\ref{dyn3d}), straightforward calculations lead, to first order in $\varepsilon$ and for $\tau>\tau_f$, to :
\begin{figure}
\includegraphics[width=1.0\textwidth]{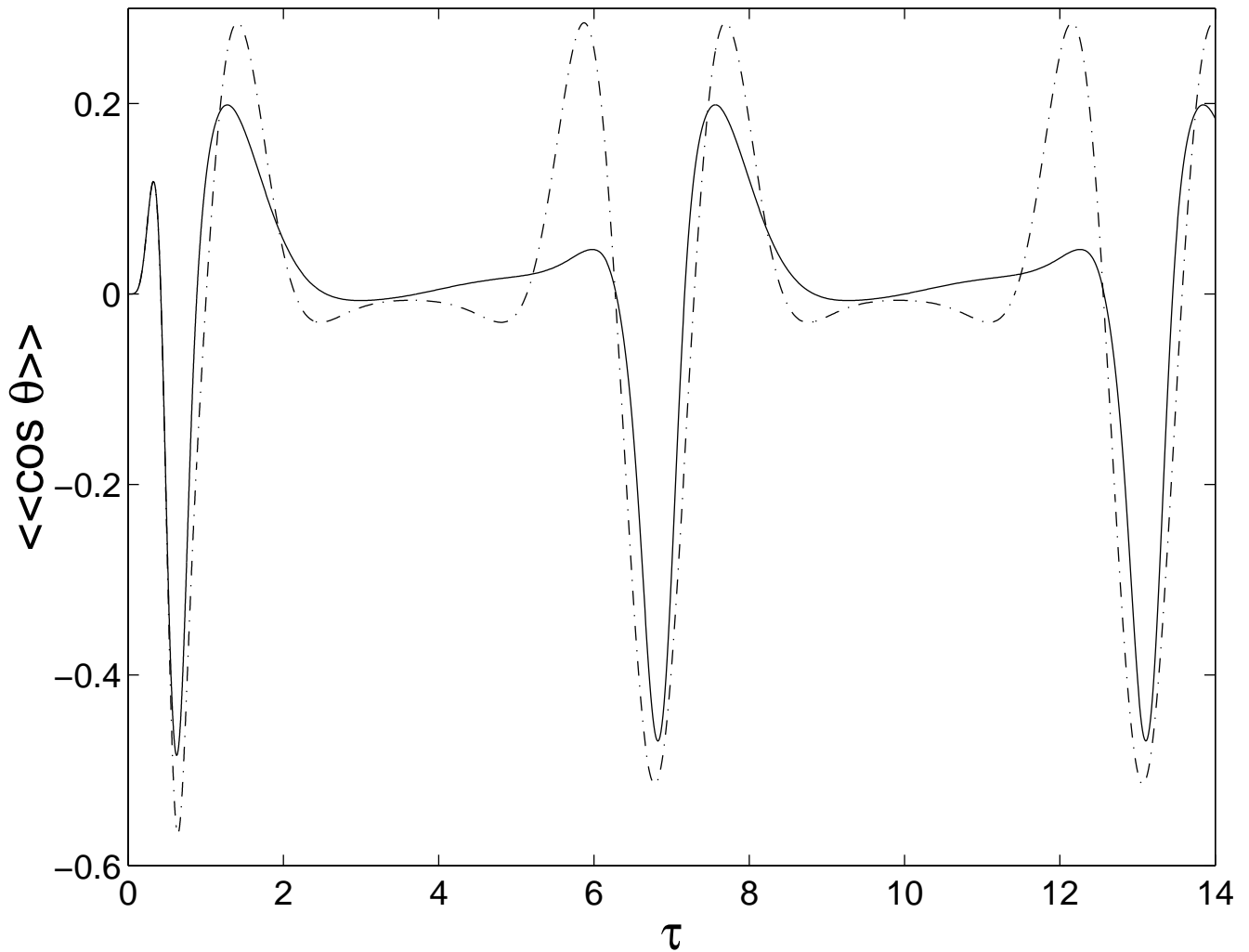}
\caption{\label{temp} Thermally averaged mean value $\ll\cos \theta \gg_{\tau}$ as a function of the adimensional time $\tau$ for the rotational temperature $T=5 \, {\textrm K}$ for $\varepsilon=0.5$, $f=2$ and $E_{0r}=70$. The pulse is on for $0\leq\tau\leq 1$. The solid line corresponds to the split-operator method and the dash-dotted line represents the result obtained from the perturbative propagator $U^I$ according to Eq.~(\ref{dyn10}).}
\end{figure}
\begin{equation} \label{dyn8}
\ll\cos\theta\gg_{\tau}=\frac{8\varepsilon}{Q}\sum_{j=1}^{+\infty} jc_j\sinh\left(\frac{Bj}{k_BT}\right)\exp\left(-\frac{Bj^2}{k_BT}\right)\int_{\tau_i}^{\tau_f}du A(u)\cos\left(2\varepsilon j[u-\tau+\tau_i]\right)+O(\varepsilon ^2)  ,
\end{equation}
where $c_j$ is defined by 
\begin{equation} \label{dyn9}
c_j=\sum_{m=-j}^{+j}c_{j\, m}  .
\end{equation}
Introducing as previously the Fourier transform $\widehat{A}$ [Eq. (\ref{dyn4b})], $\ll\cos\theta\gg_{\tau}$ can then be rewritten in the form 
\begin{eqnarray} \label{dyn10}
\ll\cos\theta\gg_{\tau}&=&\frac{8\sqrt{2\pi}\varepsilon}{Q}\sum_{j=1}^{+\infty} jc_j\sinh\left(\frac{Bj}{k_BT}\right)\exp\left(-\frac{Bj^2}{k_BT}\right)\nonumber \\
&& \times \left|\widehat{A}(2\varepsilon j)\right|\cos\left(2\varepsilon j[\tau-\tau_i]+\arg[\widehat{A}(2\varepsilon j)]\right)+O(\varepsilon ^2)  .
\end{eqnarray}
The corresponding orientation dynamics, displayed in Fig. \ref{temp} fairly reproduces the position and the value of the main extrema
of $\ll\cos\theta\gg_{\tau}$ which is the objective of this analysis. Analysing along the same lines as $T=0 \, {\textrm K}$ the role of laser parameters, one deduces that a noticeable orientation can be obtained for the following choice of numerical values : $\varepsilon=0.5$, $f=2$ and $E_{0r}=70$. In this way, an orientation efficiency of $\ll\cos\theta\gg_{\tau}\simeq 0.5$ with a duration larger than $\Delta\tau=0.3\, {\textrm {ps}}$ is achieved, which also shows the robustness with respect to temperature of this mechanism. $\Delta\tau$ is the duration over which $\ll\cos\theta\gg_{\tau}$ remains larger than $0.3$. To our knowledge, this result corresponds to one of the best reported in the literature. On the other hand, apart from these extrema, the thermally averaged $\ll\cos\theta\gg_{\tau}$ is close to the value zero, which corresponds to no orientation. More precisely, as can be clearly seen in Eq. (\ref{dyn10}), the loss of orientation between each maximum is due to the different periods $\pi/\varepsilon j$ ($j\geq 1$) of $\ll\cos\theta\gg_{\tau}$, corresponding to the different rotational frequencies of the molecule. Such an effect is basically expected since, for a freely propagating pulse, the time average of $\left\langle\cos\theta\right\rangle_{\tau}$ over a rotational period is zero \cite{dion3} 
\begin{equation}
\frac{1}{T_{{\rm rot}}}\int_t^{t+T_{{\rm rot}}} du \left\langle\cos\theta\right\rangle_{\tau}(u)=0 .
\end{equation}
\section{Summary}
This article has focused on the application of the TDUPT for studying orientation dynamics of diatomic molecules driven by pulsed laser fields.
Numerical tests have demonstrated the efficiency of the proposed procedure. A basic feature and advantage of the method under consideration,
where the small perturbation parameter is the short pulse duration, is that it allows,  through an analytical description of the molecular dynamics,
for a thorough interpretation of intense laser induced orientation. 

Until now molecular orientation with short laser pulses had been
only envisaged with non-zero time-average of the electric field  as the standard sudden impact  propagator predicts no post-pulse orientation
when this time-average vanishes.
Since a free propagating electromagnetic wave must posses
a zero time-averaged electric field, 
 half-cycle pulses  have been used.
They are composed of a short intense  pulse with non-zero time-average (which induces the molecular orientation), and a long weak tail (which is neglected).
In this paper we considered symmetrical short laser pulses with zero time-average which are easier 
to produce experimentally. 
We constructed a perturbative propagator that enables us
to elucidate the post-pulse orientation dynamics. For zero time averaged short pulses and low rotational temperatures, we have shown that the orientation may be significant depending on 
 two leading parameters.
On the one hand there is the adimensional parameter $\varepsilon \equiv B \delta$ where $B$ is the rotational constant and $\delta$ the pulse duration, and on the other hand, $\widehat A(2\varepsilon)$, the Fourier transform of the pulse area evaluated at the rescaled frequency between the first two rotational levels.
The orientation is proportional to $|\widehat A(2\varepsilon)|$ which for a prototype of symmetrical pulses is itself proportional to the peak amplitude and inversely proportional to the frequency.
Finally, we investigated the effect of temperature and showed that a good orientation  with a shorter duration can be achieved for finite but low temperatures (e.g., up to $5 \, {\textrm K}$ in the case of LiCl) by an adequate choice of these pulse parameters.

\end{document}